%% file: paper.tex
\documentclass[draft,a4paper]{article}
\usepackage{prelim2e}
\usepackage{sty/fixme}

\usepackage{ucs}
\usepackage[utf8x]{inputenc}
\usepackage[T1]{fontenc}
\usepackage[english]{babel}
\usepackage[spaces,hyphens]{url}
\usepackage{ifpdf}

\usepackage[printonlyused,smaller]{acronym}
\usepackage{booktabs}
\usepackage{color}
\usepackage{sty/cpp}
\usepackage{enumitem}
\usepackage{float}%
  \floatstyle{ruled}%
  \newfloat{Algorithm}{tp}{lsa}%
\usepackage[final]{graphicx}
\usepackage{sty/listings}
\usepackage{sty/mathitbf}
\usepackage{relsize}
\usepackage{rotating}
\usepackage{subfigure}

\usepackage{hyperref}

\usepackage{authblk}
\title{
VM-MAD: a cloud/cluster software for service-oriented academic environments
}
\author[2]{Tyanko~Aleksiev}
\author[1]{Simon~Barkow}
\author[3]{Peter~Kunszt}
\author[2]{Sergio~Maffioletti}
\author[2]{Riccardo~Murri}
\author[1]{Christian~Panse}
\PrerenderUnicode{üä}
\affil[1]{%
  Functional Genomics Center Zürich\\
  ETH Zürich / Universität Zürich\\
  Winterthurerstrasse 190, CH-8006 Zürich, Switzerland\\
  e-mail:~\url{cp@fgcz.ethz.ch}, \url{simon.barkow@fgcz.uzh.ch}
}
\affil[2]{%
  Grid Computing Competence Center\\
  Universität Zürich\\
  Winterthurerstrasse 190, CH-8006 Zürich, Switzerland\\
  e-mail:~\url{tyanko.alexiev@gmail.com}, \url{sergio.maffioletti@gc3.uzh.ch}, \url{riccardo.murri@gmail.com}
}
\affil[3]{%
  SystemsX, ETH Z\"urich\\
  Clausiusstrasse~45, CH-8092 Z\"urich, Switzerland\\
  e-mail:~\url{peter.kunszt@systemsx.ch}
}
\date{Jan.~10, 2013}

\usepackage{sty/fixme}

\begin{document}
\maketitle

\begin{abstract}
The availability of powerful computing hardware in ~\acs{IaaS}
clouds makes cloud computing attractive also for comput\-ational workloads that
were up to now almost exclusively run on \acs{HPC} clusters.

In this paper we present the \acs{VM-MAD} \emph{Orchestrator}
software: an open source framework for cloudbursting Linux-based
\acs{HPC} clusters into \acs{IaaS} clouds but also computational
grids. The \emph{Orchestrator} is completely modular, allowing
flexible configurations of cloudbursting policies.  It can be used
with any batch system or cloud infra\-structure, dynamically extending
the cluster when needed. A distinctive feature of our framework is
that the policies can be tested and tuned in a simulation mode based
on historical or synthetic cluster accounting data.

In the paper we also describe how the \acs{VM-MAD} \emph{Orchestrator}
was used in a production environment at the \acl{FGCZ} to speed up the
analysis of mass spectrometry-based protein data by cloudbursting to
the Amazon \acl{EC2}. The advantages of this hybrid system are shown
with a large evaluation run using about hundred large \ac{EC2} nodes.
\end{abstract}

\section{Introduction}
\label{sec:intro}

Recent years have seen great advances in virtualization technologies,
to the point that it is now possible to run computationally-heavy
workloads on completely virtualized infrastructures.  Starting with
Amazon \ac{EC2}, commodity on-demand virtualized compute
infrastructures\footnote{Commonly referred to as ``\ac{IaaS}
  clouds''.} have become affordable to anyone. They include
virtualized compute and storage hardware, dedicated networking and a
software stack entirely under control of the end-user.

Therefore, the use of virtualized computational infrastructures has
become very appealing to smaller research groups: it is now possible
to access large computational resources without the need to buy and
maintain a corresponding hardware infrastructure.

Today, emerging computational disciplines~(e.g., Bio\-infor\-matics,
Medical informatics)~ are showing usage patterns that do not fit well
in the traditional \ac{HPC} model of few individual jobs making use of
the entire infrastructure through massively parallel programming.
Their model is to submit a very large number of small jobs in bursts
to analyze the relevant data, and then post-process the results to get
a statistical overview or model prediction. Their need for
computational resources in terms of CPU hours is similar to the
massively parallel HPC use-cases but without the need for low-latency
networks for MPI communication.  \ac{HPC} resource providers, who need
to support such user communities with transient ``peak'' workloads,
cannot afford to plan the infrastructure for peak usage, as it would
be underutilized for most of the time. At the same time, they do not
want to see a negative impact on the traditional HPC cluster users
either. Therefore, exploitation of cloudbursting to \ac{IaaS} clouds
for HPC is interesting also to small and mid-sized facilities.

The term ``cloudbursting'' describes the ability of a local
computational resource facility to dynamically add virtual machine
instances from IaaS providers to their local resource, extending it in
size elastically as needed.  Cloudbursting improves application
throughput and response time as seen by the user.  It is an efficient
technique for dynamic \ac{HPC} resources expansion and peak workload
offloading.

Cloudbursting also allows to add cluster nodes to the local resource
that extends it with new abilities to the benefit of the
users. For example, it is possible to extend the local cluster with
virtual nodes enabling Hadoop workloads, or special GPU workloads
that are not supported locally.

In this paper we present the \ac{VM-MAD} \emph{Orchestrator} software:
an open source framework for cloudbursting Linux-based \acs{HPC}
clusters into \acs{IaaS} clouds. The \ac{VM-MAD} \emph{Orchestrator}
is completely modular, allowing flexible configurations of
cloudbursting policies in the Python programming language. It can be
used with any batch-queuing cluster system or cloud infrastructure,
dynamically extending the cluster when needed.  The policies can be
tested and tuned by using the \ac{VM-MAD} \emph{Orchestrator} in
simulation mode, based on historical or synthetic cluster accounting
data.

The paper is organized as follows. We first discuss the design goals
of the \ac{VM-MAD} \emph{Orchestrator} and the architecture we devised
to implement them (Section~\ref{sec:overall}). In
section~\ref{sec:implementation} we take a more in-depth look at the
implementation and discuss how cloudbursting policies are configured
in \ac{VM-MAD}.  As a real-world use case example, we report on the
usage of the \ac{VM-MAD} \emph{Orchestrator} to run some special
ensemble jobs on the bioinformatics cluster at the \acl{FGCZ}
(Section~\ref{sec:applications}). Finally, we survey similar and
concurrently-developed solutions (Section~\ref{sec:related}) and
outline some conclusions and possible future developments
(Section~\ref{sec:conclusions}).

\subsection{List of acronyms}
\label{sec:acro}
\input{acronyms}

\section{Overall design and architecture}
\label{sec:overall}

The stated goal of the \ac{VM-MAD} project was to build a stable
software service that could be used on existing production-grade
\ac{HPC} cluster infrastructures to dynamically add computing power
during peak loads, and to automatically revert to using only local
processing facilities when the ``rush hour'' is over.  This elastic
``cloudbursting'' feature should have as little impact as
possible on the current usage patterns of \ac{HPC} clusters; ideally,
nothing should change in the \ac{HPC} users' experience but the system would
automatically launch cloud-based \acp{VM} and schedule jobs
that would otherwise not be possible or take too much time or
resources out of the cluster.

\subsection{Implementation requirements}
\label{sec:requirements}

Early in the development process, we realized that achieving these
goals entails dealing with large heterogeneity.

First of all, we would need to accommodate different batch-queuing
systems, even if we are restricting ourselves to the \ac{HPC} clusters
in use at the \ac{UZH} and the \ac{ETHZ}.  While they all share the
same workflow and interaction models, details of the submission \ac{API} vary
greatly.  This ruled out the possibility of implementing the
\ac{VM-MAD} cloudbursting software as an extension package for a
particular batch system implementation.
Instead we decided to
interact with the batch system via the available command-line tools.

The second very important consideration was the actual definition of
what is meant by ``peak load'', i.e., under what conditions the
computing power should be extended using \ac{VM} instances from the
cloud, and what kinds of jobs can be run on the elastic part of the
infrastructure. Defining peak load is a subtle matter of local policy.
Any choice of a domain-specific language would have constrained the
range of supported policies and therefore limited the applicability of the
\ac{VM-MAD} cloudbursting system.  We chose instead to allow the
definition of the local policy as a set of functions written in the Python
programming language \cite{python:website}: now a decision can
be taken on the basis of \emph{all} the data available to the
cloudbursting software (see details in Section~\ref{sec:implementation}).

Finally, the ``cloud'' ecosystem is currently very dynamic.
For our software to be useful even just in the next few years, it needs to
be able to interface to different \ac{IaaS} cloud infrastructures.

Based on these requirements we opted for a completely modular
architecture: the \ac{VM-MAD} software is a \emph{framework} for building
cloudbursting scripts, perfectly adapted to the peculiarities of each
\ac{HPC} installation. It is not a ready-made add-on for a particular
product that a systems administrator can deploy with just a few
touches to a configuration file.\footnote{%
  This is similar to the current situation for the \ac{HPC} scheduling
  softwares: low-maintenance schedulers are also limited in
  configurability and functionality, while those that are flexible
  enough to implement complex policies are often custom-built or have
  a rich and detailed configuration language.  }\space%

\subsection{Architecture overview}
\label{sec:architecture}

Our solution to the ``cloudbursting'' problem, as outlined in the
previous sections, is to build a tool which can be run as an add-on to existing batch
systems. Our ``Orchestrator'' software
implements the additional services needed to link the batch system with
an elastic \ac{IaaS} infrastructure. It' runs in the
background\footnote{It is a ``daemon'' in the UNIX terminology.} and
performs the following tasks:
\begin{enumerate}[label=\it\alph*.]
\item\label{item:1}
  Monitor the jobs queued in the batch system, and select those
  that could run on a cloud-based \acp{VM};
\item\label{item:2}
  Start and shut down \ac{VM} instances;
\item
  Add and remove \acp{VM} compute nodes to and from the cluster.
\end{enumerate}
It is very important to remark that points~{\it \ref{item:1}}
and~{\it \ref{item:2}} involve taking decisions according to
\emph{configurable policies and metrics}.  The cluster system
administrator is responsible for these policies and metrics.

\begin{figure}
  \centering
  \includegraphics[width=1.0\linewidth]{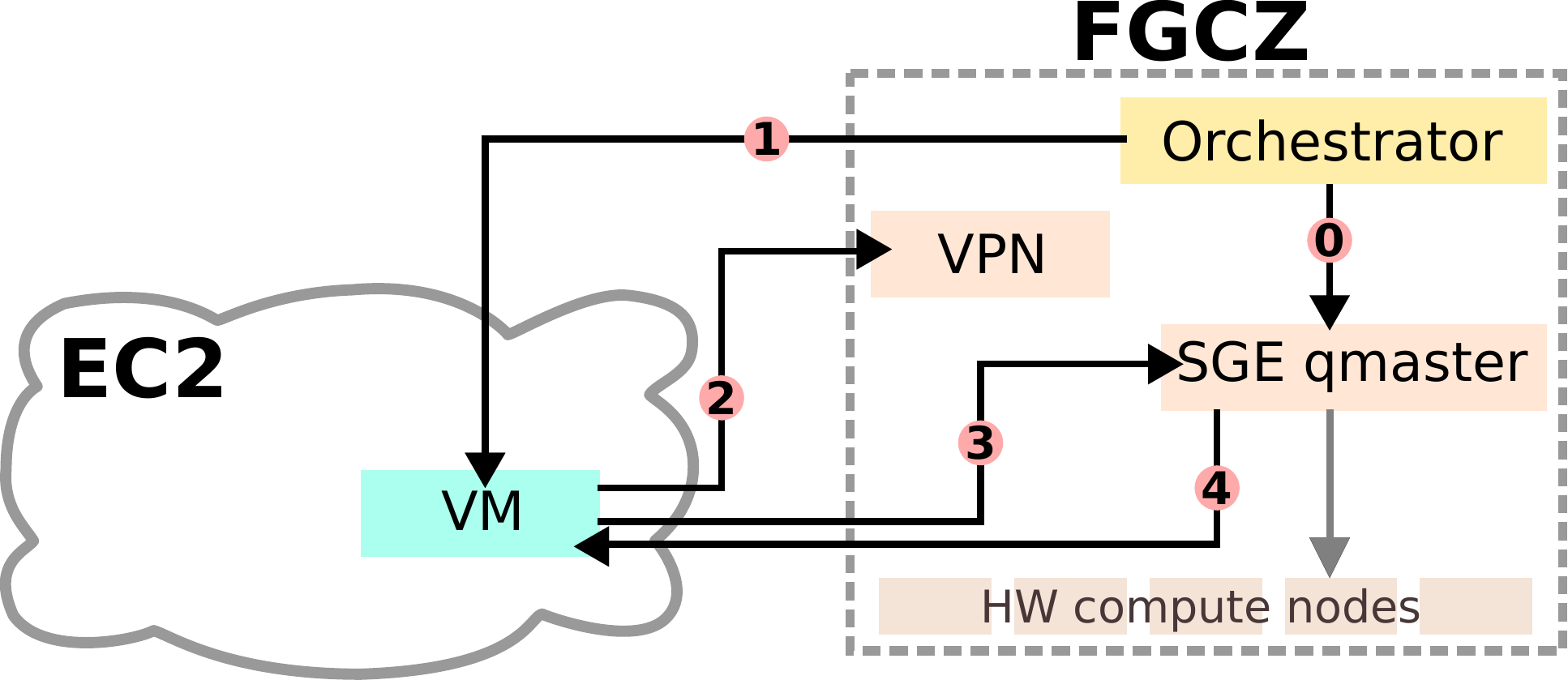}
  \caption{%
    Interaction of parts in a cloudbursting scenario.
  {\em (0)} The Orchestrator monitors the batch system state and
  determines when a new compute node is needed.
  {\em (1)} A new \ac{VM} is started.
  {\em (2)} The \ac{VM} connects back to the batch system network via \ac{VPN}.
  {\em (3)} The \ac{VM} is added to the cluster as a compute node.
  {\em (4)} The batch system can now start jobs on the \ac{VM}.
  }%
  \label{fig:cloudbursting}
\end{figure}

Figure~\ref{fig:cloudbursting} shows the interaction of the software
components involved in a cloudbursting scenario under control of a
\ac{VM-MAD} Orchestrator.
\begin{enumerate}[label=(\arabic*),start=0]
\item The Orchestrator monitors the batch system state and determines
  that ---by the local policy definition--- a new compute node is
  needed. (For example, the number of queued jobs that could be
  executed in a cloud-based \ac{VM} exceeds a certain threshold.)
\item The Orchestrator consults the cloud state and the local policy,
  and determines that the current set of cloud-based resources is
  insufficient.  It therefore contacts the cloud provider via its
  network \acs{API} and starts a new \ac{VM}.
\item The new \ac{VM} connects back to the batch system
  network via a \ac{VPN}.  This requires that the \ac{VM} image has
  been previously prepared by the cluster systems administrator: it
  should contain a the portion of the cluster execution environment
  that is necessary for running jobs destined to the cloud and the preconfigured
  \acs{VPN} software to connect back to the ``home'' network.
\item The Orchestrator adds the new \ac{VM} to the cluster as a
  compute node, reconfiguring the batch system scheduler on the
  fly. All properties of this node are registered with the scheduler
  and jobs requesting those properties can be scheduled on the new
  cloud-based nodes.
\item The batch system scheduler can now start jobs on the VM.  It
  should be noted that the Orchestrator has a passive role with
  regards to scheduling computational jobs in the cloud: all it does
  is to start new \acp{VM} that satisfy the job requirements, and lets
  the batch system scheduler use those for actually running a job.
\item When the Orchestrator detects that the amount of cloud-based
  resources exceeds the current needs (as defined by local policy), it
  shuts down the unneeded \acp{VM}.
\end{enumerate}

\section{Implementation overview}
\label{sec:implementation}

The \ac{VM-MAD} cloudbursting framework is implemented as a library
package written in the Python \cite{python:website} programming language.
The code is written in an object-oriented style; the basic components
of the framework are Python classes.

\begin{sidewaysfigure}[p]
 \fbox{
  \subfigure{%
    \includegraphics[width=0.37\textwidth]{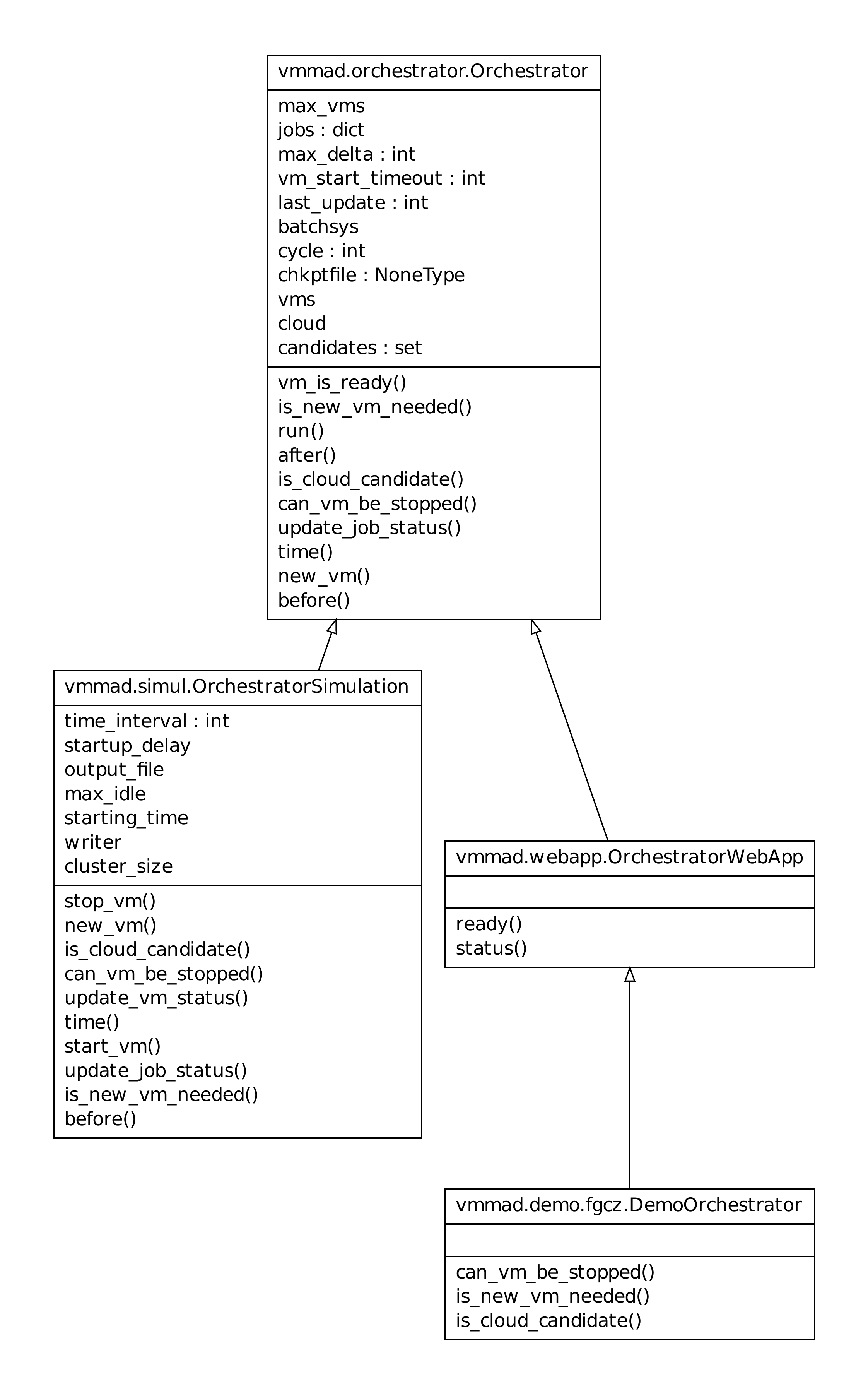}
    \label{fig:UML:vmmad.orchestrator}
  }
  \subfigure{%
      \includegraphics[width=0.57\textwidth]{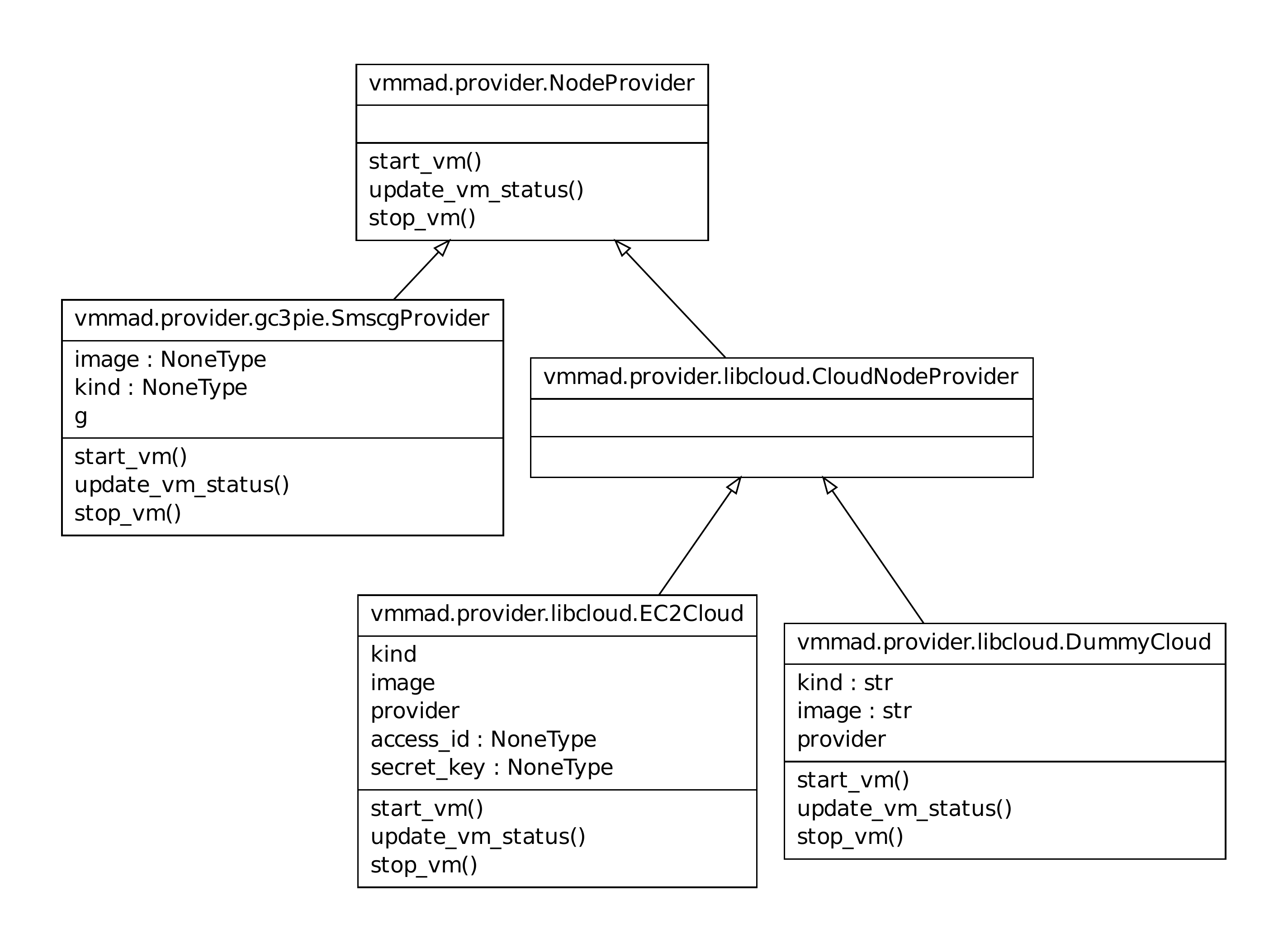}
      \label{fig:UML:vmmad.provider}
    }}
  \caption{%
    \emph{Left:}
    UML class diagram of the \emph{Orchestrator} and its derived classes.
    The root of the hierarchy is the
    \url{vmmad.orchestrator.Orchestrator} class, which implements
    the main daemon loop and the core infrastructure for the \ac{VM-MAD}
    functionality.  Two derived classes are shown: the
    \url{vmmad.simul.OrchestratorSimulation} class is used to
    simulate running the \ac{VM-MAD} software on historical accounting
    data; the \url{vmmad.demo.fgcz.DemoOrchestrator} is a an actual
    implementation of the \ac{VM-MAD} Orchestrator for use on the
    \ac{FGCZ} cluster.  Note that
    \url{vmmad.demo.fgcz.DemoOrchestrator} derives from
    \url{vmmad.orchestrator.Orchestrator} through
    \url{vmmad.webapp.OrchestratorWebApp}, which implements a web
    interface for \emph{Orchestrator} status reporting.
    \emph{Right:}
    UML class diagram of the cloud interface classes.
    The root class \url{vmmad.provider.NodeProvider} defines the programming
    interface to which other classes must conform.
    Classes in the \url{vmmad.provider.libcloud} package implement interfaces to
    different \acs{IaaS} cloud stacks using the Apache LibCloud
    library.
    The \url{vmmad.provider.gc3pie.SmscgProvider} draws nodes from
    clusters participating in the \acs{SMSCG} computational grid
    infrastructure; it is an example of how \ac{VM-MAD} can be
    interfaced to non-cloud infrastructures.
  }
  \label{fig:UML}
\end{sidewaysfigure}

The \emph{Orchestrator} object is the core of the framework: it
implements the main loop and performs housekeeping of the shared data
structures.  The \emph{Orchestrator} is a singleton: only
one single instance should be monitoring a given batch system.  An
\emph{Orchestrator} instance must be adapted to the cluster setup by
initializing it with a \emph{SchedInfo} and a
\emph{Provider} instance.

\emph{SchedInfo} objects are responsible for interacting with the
batch system scheduler, especially for gathering information
about the running/queued jobs and the available compute nodes.
New batch-queuing systems can be supported by creating an appropriate
\emph{SchedInfo} subclass.

\emph{Provider} objects are responsible for interacting with a remote
\ac{IaaS} cloud system and starting/stopping virtual machines.

The cloudbursting policy is defined by subclassing the
\emph{Orchestrator} object and overriding well-defined methods that
decide whether a job is a candidate for cloud execution, or what type
of virtual machine should be started.

It should be noted that this simple component architecture allows a
great deal of flexibility: for instance, a \emph{Provider} instance needs
not interface to a cloud provider, but can also request nodes from a
peer cluster or Grid infrastructure.\footnote{%
  This has actually been done in the course of the \ac{VM-MAD}
  benchmarks, by starting \acp{VM} \cite{apppot,apppot:website} on the \ac{SMSCG}
  \cite{smscg:website} computational grid infrastructure.
}\space%
Likewise, the \emph{SchedInfo} component does not need to read
information from a live batch system: the standard \ac{VM-MAD}
software distribution includes components for replaying job
information from a batch system accounting file, which can be used for
simulating the effect of cloudbursting policies over historical data,
see section \ref{sec:simulator}. It also includes components to
generate random workloads to be used for testing of the system and
available infrastructure.


\subsection{Policy definition}
\label{sec:policy}

``Orchestrator policies'' are criteria that govern
decisions on whether:
\begin{enumerate}
\item a given job can run on cloud-based virtualized hardware;
\item a new \acp{VM} should be started to extend the current
  virtualized computational resource pool;
\item a running \acp{VM} should be stopped, shrinking the current virtualized resource pool.
\end{enumerate}
For each of these decisions, a method is provided in the
\emph{Orchestrator} class that should return a \emph{True/False} value
based on the evaluation of available data.  Systems administrators
should override the default implementation to implement their chosen
criteria.

\subsubsection{Example: Policy on jobs eligible to run on virtualized hardware}
\label{sec:policy-jobs}

The decision whether a certain job can run on cloud-based resources is taken by
the \lstinline"is_cloud"\-\lstinline"_candidate" method.  This method is called
once for each new job that appears in the batch system queues and
returns \lstinline"True" if that job is eligible for cloudbursting.
The default implementation always returns \lstinline"False", so that
no job accidentally triggers the spawning of cloud-based \acp{VM}.

For example, the following code would implement a policy where only
jobs that have been submitted to a special ``cloud'' queue trigger
cloudbursting of compute resources:
\begin{python}
def is_cloud_candidate(self, job):
  return (job.queue == 'cloud.q')
\end{python}

The \emph{job} record passed as argument to the
\lstinline"is_cloud"\-\lstinline"_candidate" method contains all the information
that the batch system scheduler provides via its queue-listing command
(e.g., \texttt{qstat} on Sun/Oracle Grid Engine). %

\subsubsection{Example: Policy on starting new compute resources}
\label{sec:policy-start}

The decision on whether new cloud-based resources should be requested is
taken by the
\lstinline"is_new" \lstinline"_vm_needed" method. This method is
called at each iteration of the Orchestrator's main loop. It has
access to all the internal data structures, in particular the list of
jobs eligible to run on cloud-based hardware (\lstinline"self.candidates")
and the list of cloud-based \acp{VM} that have already been started by
the Orchestrator (\lstinline"self.vms").  By default, this method always
returns \lstinline"False", so that cloud-based \acp{VM} are never
spawned; this is a safety measure to avoid that non-configured
Orchestrators start spawning \acp{VM}: since usage of cloud-based
resources usually comes at a cost, it is entirely the administrator's task
to decide when and how to initiate cloudbursting.

For example, the following code implements a policy where new
cloud-based \acp{VM} are started if the number of queued candidate
jobs is greater than double the number of \acp{VM} required to run them:
\begin{python}
def is_new_vm_needed(self):
  if len(self.candidates) > 2*len(self.vms):
    return True
  else:
    return False
\end{python}

\subsubsection{Example: Policy on stopping cloud-based compute resources}
\label{sec:policy-stop}

At every iteration of the Orchestrator's main loop, a decision will
also be taken on whether an idle \ac{VM} (i.e., one that is not
currently running any job) should be stopped.  Since booting a
cloud-based \ac{VM} can take up to a few minutes' time, and many cloud
infrastructure bill usage in hourly increments, it makes sense to try
to re-use already-started \acp{VM} instead of starting new ones.  The
\lstinline"can_vm_be_stopped" method is there exactly for this
purpose: change the default Orchestrator behavior, which is to stop a
\ac{VM} as soon as it turns idle.

For example, the following code implements a policy where a \ac{VM} is
allowed to be idle for 10 minutes before it is stopped by the Orchestrator:
\begin{python}
def can_vm_be_stopped(self, vm):
  TIMEOUT = 10*60 # 10 minutes
  if vm.last_idle > TIMEOUT:
    return True
  else:
    return False
\end{python}

\section{Application and Testing}
\label{sec:applications}
For testing we have chosen an area we have a lot of expertise in. Identifying proteins in a biological sample with the help of large computer systems is a common application in the life sciences which behaviour is well studied so that we have enough experience with all parameters and configuration details, e.g. memory consumption, input-output, and stability.

\subsection{Test case: Analyzing mass spectrometric related protein data}
The processing of mass spectrometry data can be challenging as it
involves several computationally demanding algorithmic steps. Examples are
the peptide spectrum assignment of mass spectrometry data to
identify proteins in a biological sample, as well as the detection and
identification of post-translational modifications of proteins. Both tasks
can be computed simultaneously and can easily occupy hundreds of \acp{CPU} for several days.

With every new mass spectrometer, the amount of measured data
increases and the local computing infrastructures would need to be
extended accordingly.  However, these computing resources are only
needed for a short period of time. The computation demand varies
widely with the actual measurement type and the corresponding data set
size. To be able to meet also larger use-cases, the available local
cluster would need to be very large and powerful, but then it would be
mostly under-utilized.  Therefore such large use-cases are not
feasible as currently the capacity cannot be extended on demand.

Large-scale so-called ``shotgun experiments'' with complex samples from, e.g., human or fruit fly involve about ten thousand proteins. The peptide spectrum matches for our test were computed with the SEQUEST and OMSSA search algorithm
\cite{eng:1994,omssa}. To benchmark the \acs{VM-MAD} Orchestrator, we have
run the search on both the local \ac{HPC} cluster facility at the
\ac{UZH}, as well as on the Amazon \ac{EC2} Cloud
computing resources in the Amazon region US-East. To avoid denial of service like failures on the cloud system, e.g., during file
server and authentication operations we started our virtual machines
in a staggered manner with a delay of 60 seconds. In order to avoid problems like hanging processes, that might be caused by a high latency of the network connection (e.g. for accessing a network filesystem), each of our compute jobs is responsible for
dealing with its own input and output data.

\subsection{Effectiveness: Benchmark of a real world data set}
\begin{sidewaysfigure}[p]
\centering
\includegraphics[width=0.99\linewidth]{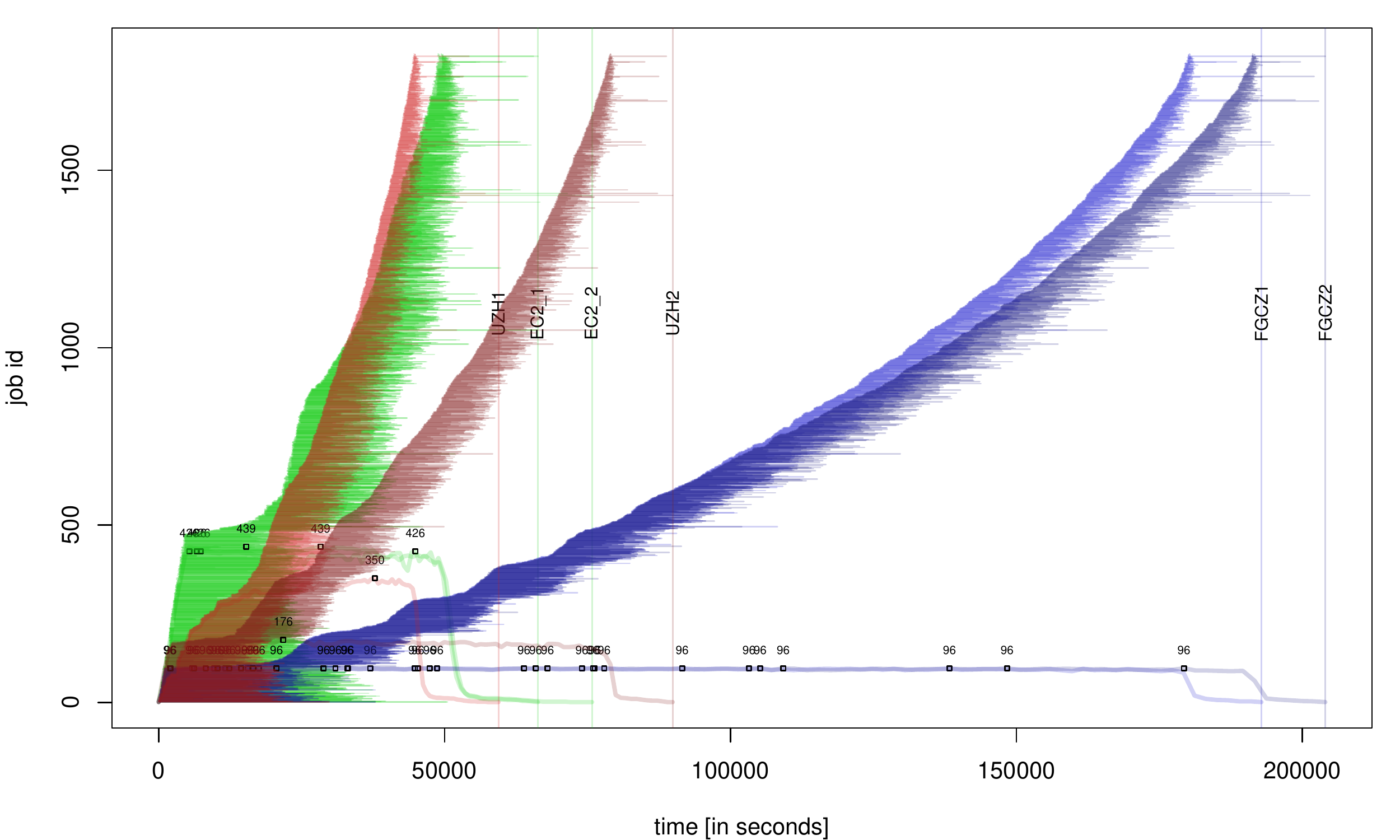}
\caption{\label{fig:cloudUtil} Overview of the benchmark. On the utilization
graph each horizontal line indicates the start and end of each job. The graph
shows that the lines for the two jobs runs on the cloud (green) have almost the
same length (we cannot distinguish 2 green branches) while the running times on
the cluster nodes (red and blue) differ more significantly (the two repetitions
are clearly distinguishable). This can be explained by the variable queue status
of the cluster nodes because of other users using the cluster at the same time.
Also, it takes much longer to run through all jobs on the limited \ac{FGCZ} cluster
(blue).
The  lines in the lower part of the graphic show the total number of
concurrently running jobs. The squares on those lines indicate the maxima on the
respective system.}
\end{sidewaysfigure}

As a test data set we used a large scale proteomics \emph{Drosophila} (fruit
fly) experiment \cite{brunner:2007} consisting of 1800 (LC)-MS/MS
runs, having a peptide mass window of 3~Dalton, 8474960~tandem mass
spectra, 498000 redundant peptides, 72281 distinct peptides, and 9124~proteins.
The data volume is approximately 0.3TB split into 1800 jobs. The
whole experiment data and the graphics are included in the {\em cloudUtil} R-package \cite{cloudUtil}.
In our benchmark we compare three compute systems: the small cluster at
the \ac{FGCZ} consisting of around 100 \acp{CPU}, a larger system as part of the
Schroedinger cluster of the \ac{UZH}, and a virtual
cluster on Amazon \ac{EC2}. For benchmarking we recorded network bandwidth,
\ac{CPU} performance (compute time) and robustness on all systems.  An
overview of the experiment and a relative comparison of each compute
job can be seen in the utilization plot on Figure~\ref{fig:cloudUtil}.

\begin{figure}
\centering
\includegraphics[width=0.99\linewidth]{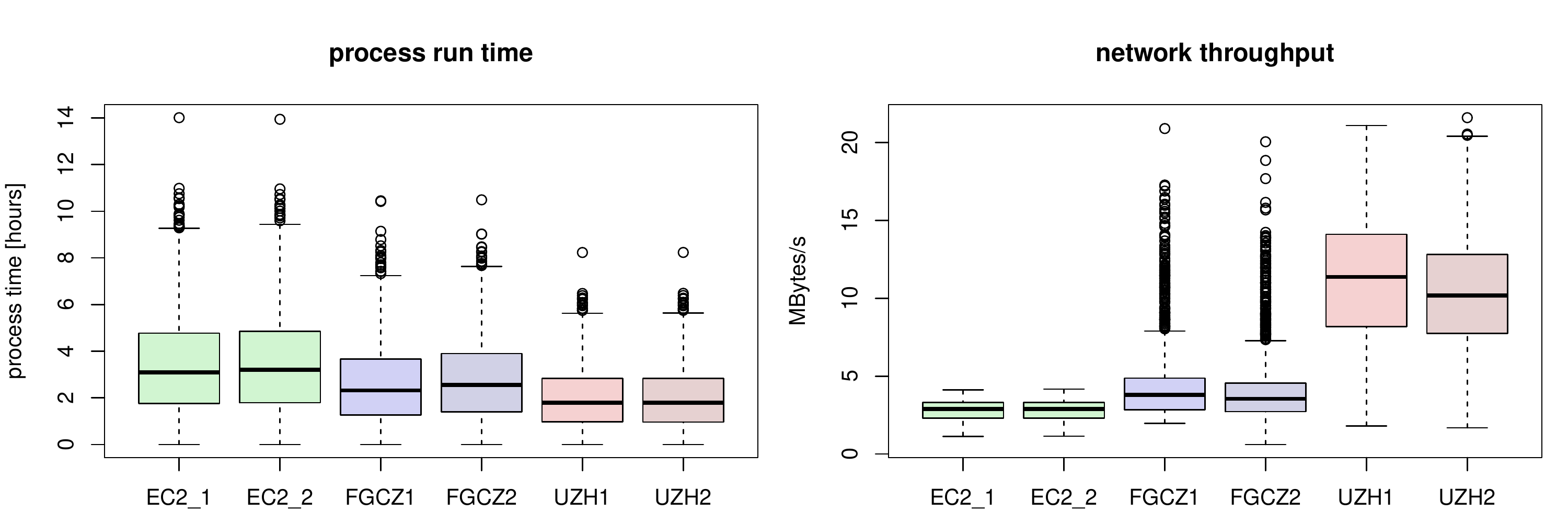}
\caption{\label{fig:cloudUtilBoxplot}Comparison run time and copy input I/O -- The box plots \cite{visualizing} display the job run time distributions of the two repetitions of all three compute systems (left) and the copy I/O network throughput (right).}
\end{figure}

The box plots \cite{visualizing} in Figure~\ref{fig:cloudUtilBoxplot}
show a comparison of the run time and the network throughput on three
compute systems having two repetitions. One \ac{ECU}
provides the equivalent \ac{CPU} capacity of a 1.0-1.2 GHz 2007 Opteron or
2007 Xeon processor. The \ac{FGCZ} cluster is based on Intel Xeon \ac{CPU} E5450
3.0GHz and the \ac{UZH} cluster is based on Intel Xeon CPU 5500.

\subsection{Simulation of LRMS accounting data}
\label{sec:simulator}
\begin{sidewaysfigure}[p]
  \centering
  \includegraphics[width=0.99\linewidth]{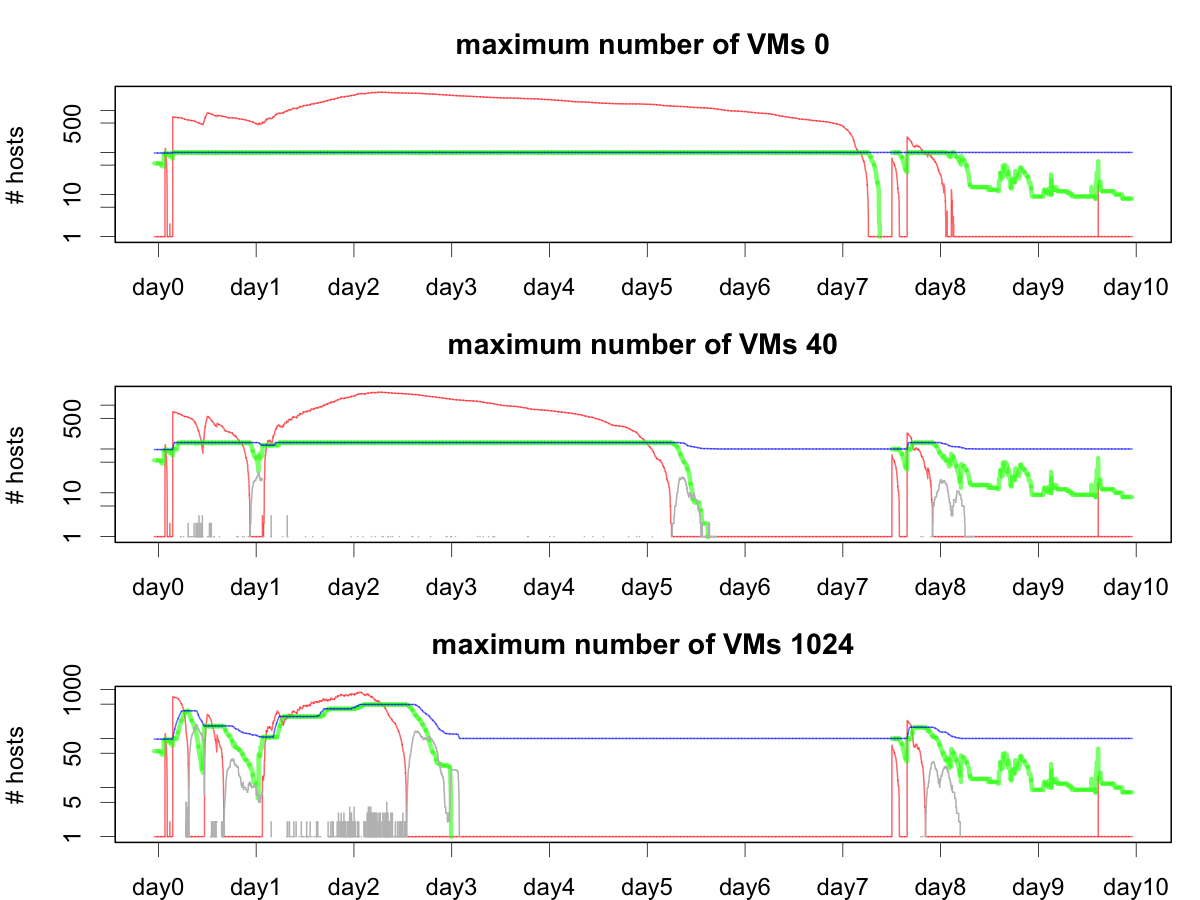}
  \caption{%
    The graphics show the simulation of 10 days of \ac{FGCZ} batch cluster accounting
    data. The vertical axis showing the number of hosts/jobs is $\log_{10}$-scaled.
    The colored lines have the following meaning:
  {\em (red)} pending jobs;
  {\em (green)} running jobs;
  {\em (blue)} available nodes to the cluster (100 plus \acp{VM});
  {\em (grey)} idle \acp{VM}.
The upper simulation run corresponds to the \ac{FGCZ} setup of 100
\acp{CPU}. For the simulation depicted in the
lower graphic we added on demand up to 40 and 1024 \acp{VM}.
For the computation we have submitted the described proteomics data set.
It can be seen that with increasing number of \acp{VM} the overall compute
time can be reduced to several days.
}%
  \label{fig:cloudsimulation}
\end{sidewaysfigure}

For demonstrating the effectiveness of the Orchestrator software and to
study the behavior of the Orchestrator policies for different hardware
scenarios (i.e. number of nodes) and different accounting data we have
implemented a simulation mode policy. If this policy is used the
\emph{Orchestrator} takes the batch-system accounting information and
the cloud configuration file as input. The accounting file of the LRMS
contains the ordered start time-stamps of every compute job and its
corresponding run time. The configuration parameters are the time step
argument (in seconds), the start time of the simulation, the maximum
number of available hosts. Instead of orchestrating real nodes the
Orchestrator writes all decisions about starting or stopping virtual
machines to an output file. The visualization in
Figure~\ref{fig:cloudsimulation} shows the simulated state of the LRMS
queue and the status of the \acp{VM} over time for different simulation
runs. In particular, the plot on the bottom of
Figure~\ref{fig:cloudsimulation} displays the output of the following
command line:
\begin{codexmp}
simul.py --time-interval 30 --start-time '2008-12-16 02:13:50 CET' \
        --max-vms 512 --cluster-size 100 --csv-file accounting.csv
\end{codexmp}

The simulation mode can also be used for determining the optimal
number of compute nodes for a given task; see. e.g.,
Figure~\ref{fig:cloudsimulation}.

\section{Related work}
\label{sec:related}

Cloudbursting, as a compute model where local resources elastically
allocate cloud instances for improving application throughput/response
time, was first proposed by Amazon’s Jeff Barr \cite{Barr:cloudbursting}.
There is a variety of different mechanisms for cloudbursting an on-premise
computational cluster to an external cloud provider: the most common
derives from the HTCondor glide-in model \cite{condor:glidein} that is used to
add a machine running on an external provider to an existing HTCondor
pool. HTCondor glide-in configures a remote resource such that it reports to and joins the local HTCondor pool. This is the technology used for example by CycleComputing.com.

Inspired by this model, workload management systems that do support
cloudbursting, like Sun/Oracle Grid Engine \cite{sge:website}, Moab
\cite{moab:website}, or the HTCondor CloudScheduler
\cite{cloudscheduler.org}, allow to start a pre-configured virtual
instance, that can reside on an external cloud provider, and let it
join the pool of resources they control. While \ac{VM-MAD} takes an
open approach in providing cloudbursting capabilities that could be
adapted to virtually any workload management system thanks to its
plug-in based approach, Grid Engine and Moab do provide a
vendor-specific solution based on policies and configurations that
cannot be applied nor ported to other similar systems.

Another approach in supporting cloudbursting is
provided by the Multi-Cluster \cite{lsf-multicluster:website} solution
from \acs{IBM}'s Platform Computing. An existing on-premise
\ac{LSF}-controlled cluster could be extended by starting an entire
\ac{LSF} cluster on a cloud provider and use Multi-Cluster to federate them
The main limitation of these approaches is the lack of an automatic
system to start and control an \ac{LSF} cluster on a cloud provider.

In terms of cloudbursting out of applications, Software-as-a-Service
solutions make use of \ac{IaaS} clouds to assure their workloads are scaled
properly. An example in the life science domain is the Galaxy CloudMan
project \cite{galaxy-cloudman,galaxy-deploy}. Here the Galaxy portal
makes use of cloud resources to extend the support for selected
computational workflows.
The CloudMan system, that is tightly coupled to the Galaxy portal, provides a
pre-selected set of tools and services as well as the possibility of
deploying own software tools and integrate them through a web interface.
While Galaxy targets the sequencing community, with ProteoCloud
\cite{Muth:2013ew}, there exists also a cloud computing pipeline for
proteomics applications but it does not feature automatic cloudburst
functionality.

In contrast to the these specialized frameworks, our approach is not
limited to life-science applications.
Any community-specific portal that is already capable of using
on-premise computational clusters could seamlessly profit from clouds
by deploying the \ac{VM-MAD} \emph{Orchestrator}.

An example of cloudbursting from an on-premise cloud infrastructure to
an external provider is brought by the Seagull project
\cite{Guo:2012:SIC:2342821.2342854}.
Seagull dynamically decides which running applications can be moved
from the on-premise cloud infrastructure to the configured external
provider, using an Intelligent Placement module based on a placement
algorithm that picks those applications to move that free up the most
units of local resources relative to their cost of running in the
cloud using a pre-defined cost function.
To reduce cloudbursting latency (due to the copying of the
disk image corresponding to the selected running application), Seagull
performs pre-copying by transferring an incremental snapshot of a
virtual machine’s disk-state to the cloud.
Seagull focuses on cloud-to-cloud
cloudbursting features, whereas \ac{VM-MAD} allows to cloudburst a
batch-controlled computational cluster; Seagull takes autonomous
decisions on what running applications to migrate live; on the other
hand, \ac{VM-MAD} has a simple policy module to determine whether to launch
new appliances on the connected cloud provider.

\section{Conclusions and Future Work}
\label{sec:conclusions}

In this work we have described the architecture, the implementation,
and an application use case of the \emph{Orchestrator} cloudbursting
software framework developed by the the \ac{VM-MAD} project. The
\emph{Orchestrator} allows an existing compute cluster to be
extended, burst into the cloud based on a highly configurable set of
local policies. We have successfully extended local clusters with
Amazon \ac{EC2} instances. In the future we also want to run Hadoop
applications on the virtual part of the cluster, enabling MapReduce
applications for our local users. Also, connecting to non-public
clouds, e.g., in-house OpenStack \ac{IaaS} is possible, as well as to extend to other
batch cluster systems.

The \emph{Orchestrator} is a modular framework that can be interfaced
with any batch-queuing system and \ac{IaaS} cloud infrastructure.  An
interesting consequence of this modularity is that the
\emph{Orchestrator} can also be run in \emph{simulation mode}, to
allow testing cloudbursting policies against historical accounting
data and evaluate the most cost-effective one. We will try to optimize
the usage of historical data in the future to suggest good predefined
policies to system administrators.

As a test and benchmark, we have used the \ac{VM-MAD}
\emph{Orchestrator} for re-processing a large set of proteomics data; the performance data collected show that commercial \ac{IaaS} clouds
can already deliver computational and network performance comparable
to what is offered by a small in-house cluster, and are thus suitable
for offloading peak computational workloads. We will also use the same
concept with other workloads, like computational chemistry and
structural biology, but also in the domains of geography and finance.

\bibliographystyle{plain}
\bibliography{grid.bib,proteomics.bib}

\end{document}

%% file: acronyms.tex


\begin{acronym}[MMMMMMM]
  \acro{AAI}{Authentication and Authorization Infrastructure}%
  \acro{AC}{Attribute Certificate\acroextra{ (VOMS, X.509)}}%
  \acro{API}{Application Programming Interface}%
  \acro{ARC}{Advanced Resource Connector}
  \acro{CA}{Certification Authority}%
  \acro{COMPCHEM}{Computational Chemistry}
  \acro{CPU}{Central Processing Unit}
  \acro{CSR}{Certificate Signing Request}%
  \acro{DAG}{Directed Acyclic Graph}
  \acro{DN}{Distinguished Name}%
  \acro{EC2}{Elastic Compute Cloud}%
  \acro{ECP}{Enhanced Client or Proxy}%
  \acro{ECU}{\acs{EC2} Compute Unit}%
  \acro{EGEE}{Enabling Grids for E-sciencE}%
  \acro{EGI}{European Grid Initiative}
  \acro{ETHZ}{{\em Eidgenössische Technische Hochschule Zürich}%
    \acroextra{, Swiss Federal Institute of Technology Zurich}}
  \acro{FGCZ}{Functional Genomics Center Zurich}
  \acro{FQAN}{Fully-Qualified Attribute Name\acroextra{ (VOMS)}}%
  \acro{GAMESS}{General Atomic and Molecular Electronic Structure System\acroextra{ (a Computational Chemistry application, see \cite{GAMESS:1993,GAMESS:2005})}}%
  \acro{GASuC}{Grid Applications Porting Group}
  \acro{GC3}{Grid Computing Competence Center, University of Zurich}
  \acro{GEMS}{Grid-Enabled Molecular Simulator}
  \acro{HPC}{High-Performance Computing}
  \acro{HTC}{High-Thorughput Computing}
  \acro{HTTP}{HyperText Transfer Protocol}%
  \acro{IaaS}{Infrastructure-as-a-Service}
  \acro{IBM}{International Business Machines}
  \acro{ID-WSF}{Identity Domain - Web Service Framework\acroextra{ (Shibboleth)}}
  \acro{IGI}{Italian Grid Initiative}
  \acro{IGTF}{International Grid Trust Federation}%
  \acro{IdP}{Identity Provider\acroextra{ (Shibboleth)}}%
  \acro{LSF}{Load Sharing Facility\acroextra{ (a batch-queuing system)}}
  \acro{MS}{Mass Spectrometry}
  \acro{PES}{Potential Energy Surface}
  \acro{PKI}{Private Key Infrastructure}%
  \acro{SAML2}{Security Assertion Markup Language version 2}%
  \acro{SAML}{Security Assertion Markup Language}%
  \acro{SGE}{Sun/Oracle Grid Engine\acroextra{ (a batch-queuing system)}}%
  \acro{SLCS}{Short-Lived Credential Service}%
  \acro{SMSCG}{Swiss Multi-Science Computational Grid}
  \acro{SOA}{Service-Oriented Architecture}
  \acro{SP}{Service Provider\acroextra{ (Shibboleth)}}%
  \acro{SSO}{Single Sign-On}%
  \acro{SWITCH}{Swiss Academic Network Provider}%
  \acro{SWITCHaai}{Swiss Federated AAI}\acused{SWITCHaai}%
  \acro{TERENA}{Trans-European Research and Education Networking Association}%
  \acro{UI}{User Interface}
  \acro{UML}{User-Mode Linux}
  \acro{URL}{Uniform Resource Locator}%
  \acro{UZH}{University of Zurich}
  \acro{VHO}{Virtual Home Organisation}%
  \acro{VM}{Virtual Machine}
  \acro{VM-MAD}{Virtual Machines Management and Advanced Deployment%
    \acroextra{ (the project described in this paper)}}
  \acro{VOMS}{Virtual Organisation Membership Service}%
  \acro{VO}{Virtual Organization}
  \acro{VPN}{Virtual Private Network}
  \acro{WMS}{Workload Management System}
  \acro{XACML}{eXtensible Access Control Markup Language}%
  \acro{xRSL}{Extended Resource Specification Language}
\end{acronym}
